\documentclass{jpp}

\usepackage{graphicx}  
\usepackage{dcolumn}  

\usepackage[utf8]{inputenc}
\usepackage[T1]{fontenc}
\usepackage{mathptmx}
\usepackage{amsmath,amsfonts,mathtools}
\usepackage{float}
\usepackage{xcolor}
\usepackage{hyperref}
\usepackage{listings}

\hypersetup{colorlinks=false}

\definecolor{my-gray}{gray}{0.95}
\definecolor{my-green}{HTML}{15B01A}
\definecolor{my-purple}{HTML}{9A0EEA}
\definecolor{my-yellow}{HTML}{FAC205}
\definecolor{codegreen}{rgb}{0,0.6,0}
\definecolor{codegray}{rgb}{0.5,0.5,0.5}
\definecolor{codepurple}{rgb}{0.58,0,0.82}
\definecolor{codered}{rgb}{0.95,0.02,0.03}

\lstdefinestyle{mystyle}{
    backgroundcolor=\color{my-gray},
    commentstyle=\color{codegreen},
    keywordstyle=\color{codepurple},
    numberstyle=\tiny\color{codegray},
    stringstyle=\color{codered},
    basicstyle=\ttfamily\footnotesize,
    breakatwhitespace=false,
    breaklines=true,
    captionpos=b,
    keepspaces=true,
    numbers=left,
    numbersep=5pt,
    showspaces=false,
    showstringspaces=false,
    showtabs=false,
    tabsize=2
}
\newcommand{\iotabar}{\mbox{$\,\iota\!\!$-}}

\title{The DESC Stellarator Code Suite Part III: Quasi-symmetry optimization}

\author{D. W. Dudt \aff{1}, R. Conlin \aff{1}, D. Panici \aff{1}  \and E.  Kolemen \aff{1} \corresp{\email{ekolemen@princeton.edu}}}

\affiliation{\aff{1}Princeton University, Princeton, New Jersey 08544}

\begin{document}

\maketitle

\begin{abstract}

The DESC stellarator optimization code takes advantage of advanced numerical methods to search the full parameter space much faster than conventional tools. 
Only a single equilibrium solution is needed at each optimization step thanks to automatic differentiation, which efficiently provides exact derivative information. 
A Gauss-Newton trust-region optimization method uses second-order derivative information to take large steps in parameter space and converges rapidly. 
With just-in-time compilation and GPU portability, high-dimensional stellarator optimization runs take orders of magnitude less computation time with DESC compared to other approaches. 
This paper presents the theory of the DESC fixed-boundary local optimization algorithm along with demonstrations of how to easily implement it in the code. 
Example quasi-symmetry optimizations are shown and compared to results from conventional tools. 
Three different forms of quasi-symmetry objectives are available in DESC, and their relative advantages are discussed in detail. 
In the examples presented, the triple product formulation yields the best optimization results in terms of minimized computation time and particle transport. 
This paper concludes with an explanation of how the modular code suite can be extended to accommodate other types of optimization problems. 

\end{abstract}

\maketitle

\section{Introduction}

\subsection{Motivation}

Stellarators are an attractive candidate for fusion energy generation due to their potential for steady-state operation and reduced susceptibility to major disruptions. 
Unlike axisymmetric configurations such as tokamaks, the three-dimensional (3D) magnetic geometry of stellarators unfortunately does not guarantee that all charged particles will be confined by these devices. 
Trapped particles, which do not sample full magnetic field lines, are subject to radial drifts that can cause them to cross magnetic flux surfaces and leave the plasma volume. 
This necessitates \textit{stellarator optimization}: the search for magnetic geometries that maximize the confinement of charged particles. 
There are many other physics and engineering objectives that are also important qualities of an ``optimal'' stellarator design, including the stability of the equilibrium magnetic field and the complexity of the coils needed to generate that field. 
The design space of stellarators is very large -- their 3D external magnetic fields posses about an order of magnitude more degrees of freedom than tokamaks \citep{Boozer2015} -- and this high-dimensional optimization problem is computationally challenging. 
Developing efficient methods to search this large parameter space and find optimal solutions is essential for the continued improvement of stellarator performance towards practical energy production. 

An ideal magnetohydrodynamic (MHD) static equilibrium magnetic field $\mathbf{B}$ and current density $\mathbf{J}$ satisfies 
\begin{subequations}
\label{eq:equilibrium}
\begin{align}
\label{eq:momentum}
\mathbf{J} \times \mathbf{B} &= \nabla p \\
\label{eq:Ampere}
\nabla \times \mathbf{B} &= \mu_0 \mathbf{J} \\
\label{eq:Gauss}
\nabla \cdot \mathbf{B} &= 0
\end{align}
\end{subequations}
throughout the plasma volume, where $\mu_{0}$ is the magnetic constant and $p$ is the pressure. 
Magnetic field lines are generally chaotic in three-dimensional geometries, but nested flux surfaces are desirable for achieving higher plasma pressures. 
This additional requirement of magnetic surfaces is given by
\begin{equation}
\label{eq:nested}
\mathbf{B} \cdot \nabla \psi = 0,
\end{equation}
where $2\pi\psi$ is the toroidal magnetic flux through a surface of constant toroidal angle $\phi$. 
Equilibrium solutions that meet the criteria of Equations \ref{eq:equilibrium} and \ref{eq:nested} are uniquely defined by the shape of their boundary surface along with the pressure and rotational transform (or plasma current density) profiles \citep{Kruskal1958}. 
Stellarator optimization seeks to find configurations with the most desirable properties in the landscape of these free variables. 

\subsection{Literature Review}

Attempts have been made to avoid searching this high dimensional space by directly constructing optimized stellarators using expansions in the distance from the magnetic axis \citep{Landreman2018,Landreman2019,Plunk2019}. 
While these methods may provide useful starting points for further optimization, they are inadequate for designing realistic stellarators with acceptable confinement near the plasma edge. 
The traditional approach to this problem is through fixed-boundary optimization, which was pioneered by Nührenberg and Zille \citep{Nuhrenberg1988}. 
The boundary shape is typically assumed to posses stellarator symmetry and parameterized in a finite double Fourier series of the form 
\begin{subequations}
\label{eq:boundary}
\begin{align}
\label{eq:Rb}
R^b(\theta,\phi) &= \sum_{n=-N}^{N} \sum_{m=0}^{M} R^{b}_{mn} \cos(m\theta-n N_{FP}\phi) \\
\label{eq:Zb}
Z^b(\theta,\phi) &= \sum_{n=-N}^{N} \sum_{m=0}^{M} Z^{b}_{mn} \sin(m\theta-n N_{FP}\phi)
\end{align}
\end{subequations}
where $\theta$ is an arbitrary poloidal angle, $(R, \phi, Z)$ are the toroidal coordinates, and $N_{FP}$ is the number of (toroidal) field periods of the device. 
For a given set of desired profiles, the Fourier coefficients $\{R^{b}_{mn}, Z^{b}_{mn}\}$ are then used as the optimization parameters to search for an equilibrium with target properties. 
A set of coils that can produce this desired plasma boundary are then found through a second optimization stage, which must consider manufacturing constraints. 
Single-stage methods that combine the equilibrium and coil optimization problems have also been explored recently and promise faster computation times \citep{giuliani_single-stage_2020}. 

Stellarator optimization is therefore cast as a conventional multidimensional optimization problem, and a plethora of algorithms are available for problems of this type. 
Derivative-free routines have been used in the past, such as Brent's method as implemented in the ROSE code \citep{rose}. 
These approaches are inexpensive per iteration, but can suffer from slow convergence and are best suited for smaller problems. 
Large scale non-linear optimization is significantly aided by gradient information, but analytic derivatives are difficult to obtain for complex objectives. 
In the Levenberg-Mardquart algorithm used by the STELLOPT code \citep{stellopt}, the required derivatives are computed through finite differences. 
The cost of this approach is that it requires numerous equilibrium evaluations, and since this scales with dimensionality it can put a practical limit on the boundary resolution $(M, N)$ available to study. 
Furthermore, finite differences approximations are very sensitive to the step size, which can make this approach inaccurate and unreliable. 
Recent progress has been made through the use of adjoint methods, as in the ALPOpt code \citep{Paul2021}. 
The adjoint approach reduces the computation burden to only two equilibrium solutions and avoids the noise of numerical derivatives \citep{Antonsen2019,Paul2020}. 
Adjoint methods are labor-intensive to implement, however, and are not applicable for all desired objectives and derivate orders. 

Another method for obtaining gradient information is \textit{automatic differentiation}: a programmatic technique that exploits the primitive operations underlying all mathematical instructions to efficiently compute exact derivatives of arbitrary (differentiable) functions to any order. 
Automatic differentiation (AD) is trivial to implement in modern programming languages and is already being used in the new stellarator coil design code FOCUSADD \citep{McGreivy2021}, but has not yet been incorporated into other optimization objectives. 
The bottleneck process for stellarator optimization is the equilibrium calculations that most physics and engineering objectives depend on, and all of the existing codes -- STELLOPT, ROSE, ALPOpt, SIMSOPT \citep{simsopt}, etc. -- primarily rely on the equilibrium solver VMEC \citep{Hirshman1983}. 
Since VMEC was written in FORTRAN without support for automatic differentiation, this approach is not available for these optimization codes and they are limited to using inaccurate or expensive gradient information. 

\subsection{Optimization with DESC}

DESC \citep{DESC} is a pseudo-spectral equilibrium solver written in Python with the latest numerical techniques, and has now developed into a full stellarator optimization code with this equilibrium solver at its core. 
Unlike the decentralized approach of other stellarator optimization suites that wrap their algorithms around calls to external codes that evaluate each objective, DESC's philosophy is to compute everything ``in-house''.  
This requires re-writing existing programs as part of the DESC package, but it allows derivative information to freely propagate throughout the code and results in significant performance improvements. 
The centralized design of DESC offers many significant advantages over existing stellarator optimization tools. 
First and foremost, the use of AD through the JAX package \citep{jax2018github} provides exact derivative information from a single equilibrium solution. 
The computation time is therefore independent of the size of the optimization space, in contrast to finite difference gradients which require a number of equilibrium calculations that scales linearly with the number of optimization parameters. 
Since AD is exact it can also improve robustness and convergence over numerical derivatives. 
In comparison to adjoint methods, AD can theoretically be used with any deterministic optimization objective, is much easier to implement, and can compute higher orders of derivatives with no additional coding. 
Furthermore, the single equilibrium required at each optimization step is found using a perturbation of the previous solution as the initial guess, which converges much faster than a ``cold start''. 
Computing equilibrium solutions has historically been the bottleneck of the stellarator optimization process, and this approach minimizes that burden. 
Altogether, DESC provides a tool to efficiently search the stellarator optimization landscape and locate configurations of interest. 

This is the final paper in a three-part series about the DESC stellarator code suite. 
Part I \citep{Panici2022} highlights the advantages of the DESC equilibrium solver through a comparison to VMEC. 
Part II \citep{Conlin2022} presents the novel perturbation and continuation methods available in DESC. 
This paper, Part III, explains the theory and practice of how these tools can be combined to solve stellarator optimization problems with DESC. 
Section \ref{sec:methods} details how the optimization problem is formulated and solved in conjunction with the equilibrium problem. 
Different metrics of quasi-symmetry are then defined as examples of optimization objectives, and code snippets are given to demonstrate how this optimization can be performed with DESC. 
In Section \ref{sec:results}, the results of a quasi-symmetry optimization problem are presented for a low-dimensional example that is easy to visualize. 
The effects of using different quasi-symmetry objectives and numerical options are compared and contrasted. 
These results are intended to demonstrate the tools that DESC provides, not to show a novel quasi-symmetric result. 
Speed comparisons are also made between DESC and STELLOPT to reveal how the computation times of both codes scale for larger problems. 
This paper only covers fixed-boundary optimization and not coil optimization, which is a natural extension and still under development. 
Global and stochastic optimization methods, such as particle swarm and genetic algorithms both used by STELLOPT and ROSE, are not discussed but could be included in a future version of the code. 

\section{Methods}
\label{sec:methods}

\subsection{Equilibrium Solver}
\label{sec:equilibrium}

As explained in a previous publication \citep{Dudt2020}, DESC solves the inverse equilibrium problem by minimizing the MHD force balance errors at a series of collocation points. 
The inverse coordinate mapping is given by $R=R(\rho,\theta,\zeta)$, $Z=Z(\rho,\theta,\zeta)$, and $\lambda=\lambda(\rho,\theta,\zeta)$, where $(R,\phi,Z)$ are the toroidal coordinates, $(\rho,\vartheta,\zeta)$ are straight field-line coordinates corresponding to the choice of toroidal angle $\zeta=\phi$, and the poloidal angle is related to the coordinate used in the boundary parameterization (\ref{eq:boundary}) through: $\vartheta=\theta+\lambda(\rho,\theta,\zeta)$. 
Nested flux surfaces are assumed, and the flux surface label is chosen to be $\rho=\sqrt{\psi/\psi_{\text{edge}}}$. 
This map between the flux and toroidal coordinate systems is discretized with global Fourier-Zernike basis sets, and the spectral coefficients are denoted as $R_{lmn}$, $Z_{lmn}$, and $\lambda_{lmn}$ ($l$, $m$, and $n$ are the radial, poloidal, and toroidal mode numbers, respectively, of the basis functions). 

Let the MHD equilibrium force balance error be defined as $\mathbf{F} = \mathbf{J} \times \mathbf{B} - \nabla p$, which has a radial and helical component: 
\begin{subequations}
\begin{align}
\label{eq:f_rho}
f_{\rho} &= \left( \sqrt{g} \left( B^{\zeta} J^{\theta} - B^{\theta} J^{\zeta} \right) - p' \right) |\nabla\rho| \\
\label{eq:f_beta}
f_{\beta} &= \sqrt{g} J^{\rho} |\mathbf{\beta}|
\end{align}
\end{subequations}
where $\mathbf{\beta} = B^\zeta \nabla\theta - B^\theta \nabla\zeta$ and $\sqrt{g}$ is the Jacobian of the $(\rho,\theta,\zeta)$ computational coordinate system. 
Evaluating each function at a series of coordinates results in a system of nonlinear equations of the form 
\begin{equation}
\mathbf{f}(\mathbf{x},\mathbf{c}) = \begin{bmatrix} f_{\rho,i} \\ f_{\beta,j} \end{bmatrix}
\end{equation}
where $i=0,\ldots,I$ and $j=0,\ldots,J$ denote the $i$\textsuperscript{th} and $j$\textsuperscript{th} collocation points for each function. 
Here $\mathbf{x}$ represents the independent variables, which are a subset of the spectral coefficients $R_{lmn}$, $Z_{lmn}$, and $\lambda_{lmn}$ such that the boundary conditions $R(\rho=1) = R^b(\theta,\phi)$ and $Z(\rho=1) = Z^b(\theta,\phi)$ are satisfied. 
The vector $\mathbf{c}$ is all of the parameters that define a unique equilibrium problem, which includes the boundary coefficients $R^{b}_{mn}$ and $Z^{b}_{mn}$, coefficients for the profiles $p(\rho)$ and $\iota(\rho)$, and the total flux through the boundary surface $\psi_{edge}$. 
An equilibrium is defined as the least-squares solution 
\begin{equation}
\label{eq:fmin}
\mathbf{x}^{*} = \text{arg min}_{\mathbf{x}} |\mathbf{f}(\mathbf{x},\mathbf{c})|^2
\end{equation}
for a fixed set of parameters $\mathbf{c}$, and is typically solved with a quasi-Newton method. 

\subsection{Optimization Approach}
\label{sec:optimization}

Stellarator optimization desires to find equilibrium solutions that also satisfy secondary objectives. 
Let the system of equations $\mathbf{g}(\mathbf{x},\mathbf{c})$ represent a set of costs that are desired to be minimized. 
Examples of quasi-symmetry objective functions are given in Section \ref{sec:quasi-symmetry}, but these could be any general engineering or physics target. 
The optimal configuration is defined as the least-squares solution 
\begin{equation}
\label{eq:gmin}
\mathbf{c}^{*} = \text{arg min}_{\mathbf{c}} |\mathbf{g}(\mathbf{x}^{*},\mathbf{c})|^2
\end{equation}
where $\mathbf{x}^{*}$ is the equilibrium solution given by (\ref{eq:fmin}) for the set of parameters $\mathbf{c}^{*}$. 

DESC employs a constrained optimization approach similar to the prediction-correction continuation method for solving systems of equations \citep{Yu1998}. 
Starting from an initial equilibrium solution $(\mathbf{x}_{k}^*,\mathbf{c}_{k})$ that it not optimal, the parameters are perturbed to a new state $(\mathbf{x}_{k+1},\mathbf{c}_{k+1}) = (\mathbf{x}_{k}^*+\Delta\mathbf{x},\mathbf{c}_{k}+\Delta\mathbf{c})$ that better satisfies the objective: $|\mathbf{g}(\mathbf{x}_{k+1},\mathbf{c}_{k+1})|^2 < |\mathbf{g}(\mathbf{x}_{k},\mathbf{c}_{k})|^2$. 
The perturbation is restricted to a subspace that maintains the equilibrium constraint, and is performed using a trust region Newton method as detailed in the remainder of this section. 
Equilibrium force balance is never satisfied exactly, however, and this ``prediction'' step may violate the equilibrium constraints slightly. 
Therefore the equilibrium problem is then re-solved with the new parameters $\mathbf{c}_{k+1}$ to get the corresponding equilibrium state $\mathbf{x}_{k+1}^*$, but this ``correction'' step should be a minor adjustment from the perturbed state $\mathbf{x}_{k+1}$. 
This process is then iterated until a set of parameters $\mathbf{c}^*$ are found that satisfy the objectives to the desired tolerance, and the Newton methods are expected to yield quadratic convergence. 
These iterations can also be included in a broader optimization loop that changes the resolution and dimensionality of the problem at each step. 

The optimization algorithm implemented in DESC can be summarized by the pseudocode in Listing \ref{lst:pseudocode}. 
Stopping criteria are determined by relative tolerances on the cost function values and step sizes, in addition to maximum numbers of iterations. 

\begin{lstlisting}[language=Python,
caption=Pseudocode outlining the DESC optimization algorithm.,
label={lst:pseudocode}][h]
while optimization stopping criteria are not met:

    perturb equilibrium solution to improve objective

    while equilibrium stopping criteria are not met:

        solve equilibrium force balance
\end{lstlisting}
%
%
%

The derivation of the optimal perturbations begins by Taylor expanding both the constraint (equilibrium) and objective functions about the current state: 
\begin{subequations}
\label{eq:Taylor}
\begin{align}
\mathbf{f}(\mathbf{x}+\Delta\mathbf{x},\mathbf{c}&+\Delta\mathbf{c}) = \mathbf{f}(\mathbf{x},\mathbf{c}) + \frac{\partial\mathbf{f}}{\partial\mathbf{x}}\Delta\mathbf{x} + \frac{\partial\mathbf{f}}{\partial\mathbf{c}}\Delta \mathbf{c} \\
&+ \frac{1}{2}\frac{\partial^2\mathbf{f}}{\partial\mathbf{x}^2}\Delta\mathbf{x}\Delta\mathbf{x}^T + \frac{1}{2}\frac{\partial^2\mathbf{f}}{\partial\mathbf{c}^2}\Delta\mathbf{c}\Delta\mathbf{c}^T + \frac{\partial^2\mathbf{f}}{\partial\mathbf{x}\partial\mathbf{c}}\Delta\mathbf{x}\Delta\mathbf{c}^T \nonumber \\
\mathbf{g}(\mathbf{x}+\Delta\mathbf{x},\mathbf{c}&+\Delta\mathbf{c}) = \mathbf{g}(\mathbf{x},\mathbf{c}) + \frac{\partial\mathbf{g}}{\partial\mathbf{x}}\Delta\mathbf{x} + \frac{\partial\mathbf{g}}{\partial\mathbf{c}}\Delta \mathbf{c} \\
&+ \frac{1}{2}\frac{\partial^2\mathbf{g}}{\partial\mathbf{x}^2}\Delta\mathbf{x}\Delta\mathbf{x}^T + \frac{1}{2}\frac{\partial^2\mathbf{g}}{\partial\mathbf{c}^2}\Delta\mathbf{c}\Delta\mathbf{c}^T + \frac{\partial^2\mathbf{g}}{\partial\mathbf{x}\partial\mathbf{c}}\Delta\mathbf{x}\Delta\mathbf{c}^T \nonumber
\end{align}
\end{subequations}
The perturbations themselves are assumed to be a combination of first and second-order terms 
\begin{subequations}
\label{eq:dx}
\begin{align}
\Delta\mathbf{x} &= \epsilon\mathbf{x}_1 + \epsilon^2\mathbf{x}_2 \\
\Delta\mathbf{c} &= \epsilon\mathbf{c}_1 + \epsilon^2\mathbf{c}_2
\end{align}
\end{subequations}
where $\epsilon \ll 1$ is some small parameter. 

This paper refers to the ``order'' of a method by the number of terms included in the perturbation expansion (\ref{eq:dx}). 
However, the terminology typically used in the optimization literature would be a degree higher than this. 
The discrepancy is because the Jacobian matrices of first derivatives such as $\frac{\partial\mathbf{g}}{\partial\mathbf{c}}$ give the Hessian matrices of second derivatives for the scalar least-squares problem in (\ref{eq:gmin}). 

\subsubsection{First Order}

Substituting (\ref{eq:dx}) into (\ref{eq:Taylor}) and collecting only the terms up to order $\epsilon$ yields: 
\begin{subequations}
\begin{align}
\label{eq:f1}
\mathbf{0} &= \mathbf{f}(\mathbf{x},\mathbf{c}) + \frac{\partial\mathbf{f}}{\partial\mathbf{x}}\epsilon\mathbf{x}_1 + \frac{\partial\mathbf{f}}{\partial\mathbf{c}}\epsilon\mathbf{c}_1 \\
\label{eq:g1}
\mathbf{0} &= \mathbf{g}(\mathbf{x},\mathbf{c}) + \frac{\partial\mathbf{g}}{\partial\mathbf{x}}\epsilon\mathbf{x}_1 + \frac{\partial\mathbf{g}}{\partial\mathbf{c}}\epsilon\mathbf{c}_1.
\end{align}
\end{subequations}
Equation (\ref{eq:f1}) can be solved for $\epsilon\mathbf{x}_1$ in terms of $\epsilon\mathbf{c}_1$
\begin{equation}
\label{eq:dx1}
\frac{\partial\mathbf{f}}{\partial\mathbf{x}}\epsilon\mathbf{x}_1 = -\mathbf{f}(\mathbf{x},\mathbf{c}) - \frac{\partial\mathbf{f}}{\partial\mathbf{c}}\epsilon\mathbf{c}_1,
\end{equation}
and substituting this expression into (\ref{eq:g1}) gives an equation for the first-order parameter perturbation: 
\begin{equation}
\label{eq:dc1}
\left[ \frac{\partial\mathbf{g}}{\partial\mathbf{x}} \left( \frac{\partial\mathbf{f}}{\partial\mathbf{x}} \right)^{-1} \frac{\partial\mathbf{f}}{\partial\mathbf{c}} - \frac{\partial\mathbf{g}}{\partial\mathbf{c}} \right] \epsilon\mathbf{c}_1 = \mathbf{g}(\mathbf{x},\mathbf{c}) - \frac{\partial\mathbf{g}}{\partial\mathbf{x}} \left( \frac{\partial\mathbf{f}}{\partial\mathbf{x}} \right)^{-1} \mathbf{f}(\mathbf{x},\mathbf{c}).
\end{equation}

Equation (\ref{eq:dc1}) is a linear system that can be solved for the optimal perturbation $\epsilon\mathbf{c}_1$ using a trust-region step, as explained in Section \ref{sec:trust-region}. 
The corresponding change to the independent variables, $\epsilon\mathbf{x}_1$, is then determined from (\ref{eq:dx1}). 
All of the Jacobian matrices are computed through automatic differentiation, and $\frac{\partial\mathbf{f}}{\partial\mathbf{x}}$ is typically already know from solving the equilibrium problem (\ref{eq:fmin}) at the previous optimization step. 
A pseudo-inverse is taken for the term $\left( \frac{\partial\mathbf{f}}{\partial\mathbf{x}} \right)^{-1}$. 

\subsubsection{Second Order}

Collecting the terms proportional to $\epsilon^2$ that were omitted from (\ref{eq:Taylor}) yields: 
\begin{subequations}
\begin{align}
\label{eq:f2}
\mathbf{0} = \frac{\partial\mathbf{f}}{\partial\mathbf{x}}\epsilon^2\mathbf{x}_2 &+ \frac{\partial\mathbf{f}}{\partial\mathbf{c}}\epsilon^2\mathbf{c}_2 + \frac{1}{2}\frac{\partial^2\mathbf{f}}{\partial\mathbf{x}^2}\epsilon^2\mathbf{x}_1\mathbf{x}_1^T \\
&+ \frac{1}{2}\frac{\partial^2\mathbf{f}}{\partial\mathbf{c}^2}\epsilon^2\mathbf{c}_1\mathbf{c}_1^T + \frac{\partial^2\mathbf{f}}{\partial\mathbf{x}\partial\mathbf{c}}\epsilon^2\mathbf{x}_1\mathbf{c}_1^T \nonumber \\
\label{eq:g2}
\mathbf{0} = \frac{\partial\mathbf{g}}{\partial\mathbf{x}}\epsilon^2\mathbf{x}_2 &+ \frac{\partial\mathbf{g}}{\partial\mathbf{c}}\epsilon^2\mathbf{c}_2 + \frac{1}{2}\frac{\partial^2\mathbf{g}}{\partial\mathbf{x}^2}\epsilon^2\mathbf{x}_1\mathbf{x}_1^T \\
&+ \frac{1}{2}\frac{\partial^2\mathbf{g}}{\partial\mathbf{c}^2}\epsilon^2\mathbf{c}_1\mathbf{c}_1^T + \frac{\partial^2\mathbf{g}}{\partial\mathbf{x}\partial\mathbf{c}}\epsilon^2\mathbf{x}_1\mathbf{c}_1^T \nonumber
\end{align}
\end{subequations}
In a similar process, equation (\ref{eq:f2}) can be solved for $\epsilon\mathbf{x}_2$ in terms of $\epsilon\mathbf{c}_1$ and $\epsilon\mathbf{c}_2$
\begin{align}
\label{eq:dx2}
\frac{\partial\mathbf{f}}{\partial\mathbf{x}} \epsilon^2\mathbf{x}_2 &= -\frac{\partial\mathbf{f}}{\partial\mathbf{c}}\epsilon^2\mathbf{c}_2 - \frac{1}{2}\frac{\partial^2\mathbf{f}}{\partial\mathbf{x}^2}\epsilon^2\mathbf{x}_1\mathbf{x}_1^T \\
&- \frac{1}{2}\frac{\partial^2\mathbf{f}}{\partial\mathbf{c}^2}\epsilon^2\mathbf{c}_1\mathbf{c}_1^T - \frac{\partial^2\mathbf{f}}{\partial\mathbf{x}\partial\mathbf{c}}\epsilon^2\mathbf{x}_1\mathbf{c}_1^T, \nonumber
\end{align}
and substituting this expression into (\ref{eq:g2}) gives an equation for the second-order parameter perturbation: 
\begin{align}
\label{eq:dc2}
\left[ \left( \frac{\partial\mathbf{g}}{\partial\mathbf{x}} \right) \right. &\left. \left( \frac{\partial\mathbf{f}}{\partial\mathbf{x}} \right)^{-1} \frac{\partial\mathbf{f}}{\partial\mathbf{c}} - \frac{\partial\mathbf{g}}{\partial\mathbf{c}} \right] \epsilon^2\mathbf{c}_2 = \\
\frac{1}{2} &\left[ \frac{\partial^2\mathbf{g}}{\partial\mathbf{x}^2} - \left( \frac{\partial\mathbf{g}}{\partial\mathbf{x}} \right) \left( \frac{\partial\mathbf{f}}{\partial\mathbf{x}} \right)^{-1} \frac{\partial^2\mathbf{f}}{\partial\mathbf{x}^2} \right] \epsilon^2\mathbf{x}_1\mathbf{x}_1^T \nonumber \\
+ \frac{1}{2} &\left[ \frac{\partial^2\mathbf{g}}{\partial\mathbf{c}^2} - \left( \frac{\partial\mathbf{g}}{\partial\mathbf{x}} \right) \left( \frac{\partial\mathbf{f}}{\partial\mathbf{x}} \right)^{-1} \frac{\partial^2\mathbf{f}}{\partial\mathbf{c}^2} \right] \epsilon^2\mathbf{c}_1\mathbf{c}_1^T \nonumber \\
+ &\left[ \frac{\partial^2\mathbf{g}}{\partial\mathbf{x}\partial\mathbf{c}} - \left( \frac{\partial\mathbf{g}}{\partial\mathbf{x}} \right) \left( \frac{\partial\mathbf{f}}{\partial\mathbf{x}} \right)^{-1} \frac{\partial^2\mathbf{f}}{\partial\mathbf{x}\partial\mathbf{c}} \right] \epsilon^2\mathbf{x}_1\mathbf{c}_1^T. \nonumber
\end{align}

Trust-region steps are again used to solve for the second order correction to the optimal perturbation, $\epsilon\mathbf{c}_2$, and then the independent variables $\epsilon\mathbf{x}_2$, using equations (\ref{eq:dc2}) and (\ref{eq:dx2}), respectively. 
One difference is that the trust-region radius used for the second order terms is restricted to be a fraction of the magnitude of the first order step. 
Note that the same matrix appears on the left-hand side of Equations (\ref{eq:dc1}) and (\ref{eq:dc2}), so once the single value decomposition is known from the first order solution it does not have to be recomputed for the second order solution. 
Also note that the Hessian matrices on the right-hand side of (\ref{eq:dc2}) are not needed in full, but can be efficiently computed with Jacobian-vector products. 
Therefore this second order approximation requires relatively little additional computation time after the first order solution is found, and the extra accuracy in representing the parameter space is usually worth the trade-off. 
This adaptive procedure helps to ensure the optimization is robust to different initial conditions. 

\subsubsection{Trust-Region Step}
\label{sec:trust-region}

Equations (\ref{eq:dx1}), (\ref{eq:dc1}), (\ref{eq:dx2}), and (\ref{eq:dc2}) are all linear systems of the form $\mathbf{A}\mathbf{y}=\mathbf{b}$. 
A trust-region method is used instead of taking the full Newton step, which improves convergence while far from the optimum \citep{Nocedal2006}.
This involves casting these equations into a problem of the form 
\begin{equation}
\text{min}_{\mathbf{y}} |\mathbf{A}\mathbf{y}-\mathbf{b}|^2, \hspace{1mm} |\mathbf{y}|\leq\delta
\end{equation}
where $\delta$ is the trust region radius. 
The solution requires a single-value decomposition of the matrix $\mathbf{A}$, but this computation time is typically insignificant compared to the automatic differentiation used to evaluate the matrix. 
The trust region radius is chosen through an adaptive procedure based on the ratio of the cost reduction predicted by the model to the actual reduction achieved by the new equilibrium solution. 
If this ratio is above $0.75$ and the trust region radius was restricting the step size, the radius is doubled for the next iteration. 
The trust region radius is reduced to a quarter of its former size if the ratio is below $0.25$ -- this includes prospective steps that fail to actually reduce the cost function, which are rejected. 

\subsection{Quasi-symmetry Objective Functions}
\label{sec:quasi-symmetry}

Charged particle confinement is not guaranteed in three-dimensional toroidal geometries, and this is typically the primary objective of stellarator optimization. 
Since neoclassical confinement calculations are historically too expensive to run within an optimization loop, quasi-symmetry is a popular proxy function for particle confinement that is much cheaper to compute. 
Quasi-symmetry ensures particles remain close to flux surfaces by approximately conserving some canonical momentum \citep{Rodriguez2020}. 
The formal definition of a quasi-symmetric magnetic field is the existence of a vector field $\mathbf{u}$ such that: 
\begin{subequations}
\begin{align}
\mathbf{B} \times \mathbf{u} &= \nabla \psi \\
\mathbf{u} \cdot \nabla B &= 0 \\
\nabla \cdot \mathbf{u} &= 0.
\end{align}
\end{subequations}
However, this definition is not amenable to computational analysis. 
Three other quasi-symmetry objective functions are implemented in DESC, and although they are theoretically equivalent, each one can have distinct numerical consequences in practice. 

\subsubsection{Boozer Coordinates}

A magnetic field is quasi-symmetric if its magnitude can be written in the form 
\begin{equation}
|\mathbf{B}| = B(\psi, M\vartheta_{B} - N\zeta_{B})
\end{equation}
where $\vartheta_B$ and $\zeta_B$ are the poloidal and toroidal angles in Boozer coordinates, respectively, and $M$, $N$ $\in \mathbb{Z}$ determine the type of quasi-symmetry.  
The optimization objective function derived from this definition is
\begin{equation}
\label{eq:fB_vec}
\mathbf{f}_{B} = \{ B_{mn} ~|~ m/n \neq M/N \}
\end{equation}
where $B_{mn}$ are the Fourier coefficients of $|\mathbf{B}|$ in Boozer coordinates on a particular surface, and the $m=n=0$ mode is always excluded. 
This corresponds to the set of non-symmetric Boozer modes, which must vanish in quasi-symmetry. 
An associated scalar, dimensionless metric is 
\begin{equation}
\label{eq:fB_hat}
\hat{f}_{B} = \frac{|\mathbf{f}_{B}|}{\sqrt{\sum_{m,n} B_{mn}^2}}
\end{equation}
which is commonly used to quantify the departure from true quasi-symmetry. 
This is the traditional definition of quasi-symmetry, and the normalized scalar has the convenient property of always being in the range $\hat{f}_{B} \in [0,1)$.  
However, the transformation to Boozer coordinates can be computationally expensive. 
DESC uses the same Boozer coordinate transformation algorithm as BOOZ\_XFORM, but was rewritten within the python code to work with automatic differentiation. 
This implementation does not take advantage of ``generalized'' Boozer coordinates and therefore assumes that an equilibrium exists \citep{Rodriguez2021-1}, but the force balance is generally never satisfied to machine precision for numerical solutions. 
Furthermore, this definition only quantifies the ``global'' deviation from quasi-symmetry on each flux surface rather than providing a ``local'' deviation throughout the surface. 

\subsubsection{Two-Term}

A magnetic field is also quasi-symmetric if the quantity 
\begin{equation}
C = \frac{\left( \mathbf{B} \times \nabla \psi \right) \cdot \nabla B}{\mathbf{B} \cdot \nabla B}
\end{equation}
is a flux function: $C = C(\psi)$. 
This definition assumes $\mathbf{J}\cdot\nabla\rho = J^\rho = 0$, which is less restrictive than the requirement of force balance used in the true Boozer form, but must hold in equilibrium by (\ref{eq:f_beta}) with the same caveat about numerical resolution. 
In Boozer coordinates, this flux function is also equivalent to 
\begin{equation}
C = \frac{M G + N I}{M \iotabar - N}.
\end{equation}
The covariant components of the magnetic field in Boozer coordinates can be computed in any flux coordinate system by 
\begin{subequations}
\begin{align}
G &= \langle B_\zeta \rangle \\
I &= \langle B_\theta \rangle
\end{align}
\end{subequations}
where the flux surface average of a quantity $Q$ is defined as 
\begin{equation}
\langle Q \rangle \equiv \frac{\int_{0}^{2\pi} \int_{0}^{2\pi} Q \sqrt{g} d\theta d\zeta}{\int_{0}^{2\pi} \int_{0}^{2\pi} \sqrt{g} d\theta d\zeta}.
\end{equation}
A more useful form that avoids computational singularities is 
\begin{equation}
\label{eq:fC}
f_{C} = \left( M \iotabar - N \right) \left( \mathbf{B} \times \nabla \psi \right) \cdot \nabla B - \left( M G + N I \right) \mathbf{B} \cdot \nabla B. 
\end{equation}
Evaluating this function at a series of collocation points on a given surface yields the vector form 
\begin{equation}
\label{eq:fC_vec}
\mathbf{f}_{C} = \{ f_{C}(\theta_{i},\zeta_{j}) ~|~ i \in [0,2\pi), j \in [0,2\pi/N_{FP}) \}.
\end{equation}
These residuals have units of $\mathrm{T^3}$; a dimensionless scalar metric is chosen to be 
\begin{equation}
\label{eq:fC_hat}
\hat{f}_{C} = \frac{\langle |f_C| \rangle}{\langle B \rangle^3}.
\end{equation}
This quasi-symmetry objective does not rely on a transformation to Boozer coordinates, but it still requires specifying the helicity of the quasi-symmetry. 
Unlike the Boozer form, this two-term definition can reveal the local quasi-symmetry errors within each surface. 

\subsubsection{Triple Product}

Finally, a magnetic field is also quasi-symmetric if the function 
\begin{equation}
\label{eq:fT}
f_{T} = \nabla \psi \times \nabla B \cdot \nabla \left( \mathbf{B} \cdot \nabla B \right)
\end{equation}
vanish throughout the region of interest. 
Similarly, this function is evaluated at a series of collocation points to yield the vector form 
\begin{equation}
\label{eq:fT_vec}
\mathbf{f}_{T} = \{ f_{T}(\theta_{i},\zeta_{j}) ~|~ i \in [0,2\pi), j \in [0,2\pi/N_{FP}) \}.
\end{equation}
This function has units of $\frac{\text{T}^4}{\text{m}^2}$; a dimensionless scalar metric is chosen to be 
\begin{equation}
\label{eq:fT_hat}
\hat{f}_{T}(\psi) = \frac{\langle R \rangle^2 \langle |f_{T}| \rangle}{\langle B \rangle^4}.
\end{equation}
The triple product form benefits from the same advantages as the two-term definition: it does not rely on a transformation to Boozer coordinates, and provides local errors rather than a flux-surface quantity. 
Furthermore, it does not assume any equilibrium conditions or require the helicity to be specified a-priori, which could be useful for certain optimization scenarios. 
This form is also more amenable to optimizing for quasi-symmetry throughout a volume rather than a single flux surface, which may not be physically achievable but could still have practical applications \citep{Garren1991-1,Garren1991-2}. 

\section{Results}
\label{sec:results}

\subsection{Code Implementation}
\label{sec:code}

DESC is designed to have an approachable user-interface through python scripts, similar to the SIMSOPT framework. 
The code snippet in Listing \ref{lst:code} demonstrates how to run DESC as it was used to generate the results in Section \ref{sec:optimization}. 
This example shows the \texttt{QuasisymmetryTwoTerm} function corresponding to $\mathbf{f}_{C}$, targeting quasi-helical symmetry. 
It is also normalized using the same denominator as (\ref{eq:fC_hat}). 
The syntax for the objective functions \texttt{QuasisymmetryBoozer} and \texttt{QuasisymmetryTripleProduct} corresponding to $\mathbf{f}_{B}$ and $\mathbf{f}_{T}$, respectively, are similar. 

The \texttt{"opt\_subspace"} argument is a transformation matrix that can be used to restrict the optimization parameters to a custom subspace. 
In this example, it is used to optimize over two boundary coefficients given in the double-angle Fourier series basis of (\ref{eq:boundary}). 
DESC uses a different but equivalent basis from this VMEC convention, and the \texttt{vmec\_boundary\_subspace} function relates them through the Ptolemy identities. 
The results presented in this paper were run with the default options in DESC, but the numerical resolution, target surfaces, stopping criteria, and all other options can be fully customized. 

Although it is not shown in this example, multiple optimization objectives can be targeted together with relative weighting between them. 
Examples of other objectives that are available in DESC include geometric quantities such as the plasma volume and aspect ratio. 
The code is designed in a modular structure so that new objective functions can easily be added to the existing framework. 

\begin{lstlisting}[language=Python,
caption=Example code to run quasi-helical symmetry optimization using the two-term objective function.,
label={lst:code}][h]
from desc import set_device
set_device("gpu")  # if running on a GPU

from desc.equilibrium import Equilibrium
from desc.objectives import (
    ObjectiveFunction,
    QuasisymmetryTwoTerm,
    get_fixed_boundary_constraints,
)
from desc.vmec_utils import vmec_boundary_subspace

# load an initial equilibrium solution
eq = Equilibrium.load("path/to/equilibrium.h5")

QH = (1, eq.NFP)  # type of quasi-symmetry

# objective function to optimize
objective = ObjectiveFunction(
    QuasisymmetryTwoTerm(helicity=QH, norm=True),
    get_fixed_boundary_constraints(),
)

# optimization variables
perturb_options = {
    "dRb": True,  # optimize R boundary mode RBC(2,1)
    "dZb": True,  # optimize Z boundary mode ZBS(2,1)
    "opt_subspace": vmec_boundary_subspace(
        eq, RBC=[2, 1], ZBS=[2, 1]
    ),
}

# run optimization
eq.optimize(objective, perturb_options=perturb_options)
\end{lstlisting}
%
%
%
%
%
%

\subsection{Quasi-Symmetry Optimization}
\label{sec:optimization}

This section demonstrates the capability of DESC to optimize stellarator equilibria, using quasi-symmetry as an example objective function. 
As the initial state, the $m=1$, $n=2$ boundary modes for both $R^{b}$ and $Z^{b}$ in (\ref{eq:boundary}) of a quasi-helically symmetric STELLOPT benchmark solution \citep{LandremanGitHub} were modified to degrade the quality of quasi-symmetry. 
The equilibrium was then optimized for quasi-symmetry on the last closed flux surface ($\rho=1$) in this two-dimensional parameter space with all other boundary modes and profiles held fixed at the benchmark solution values. 
The optimization was performed using each of the three objective functions described previously, and with both first and second-order optimization methods. 
This simple design space was chosen for comparison to a previous study of this problem \citep{Rodriguez2022}, and it lends itself well to visualization. 
Note that in the example used here (in contrast to the previous study), quasi-symmetry is only being targeted on a single flux surface (instead of multiple surfaces) and the rotational transform profile (instead of the current profile) is held fixed during optimization. 

\begin{figure}
\centering
\includegraphics[width=0.6\textwidth]{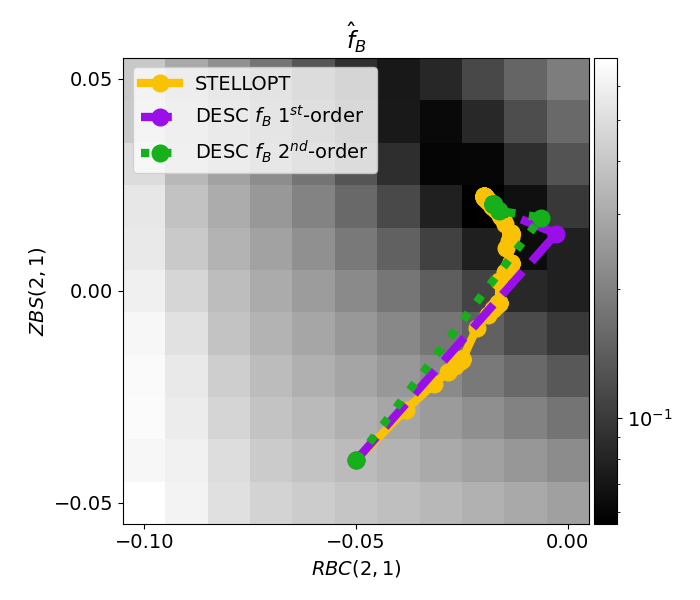}
\caption{Quasi-symmetry optimization paths using the Boozer coordinate objective in DESC. The STELLOPT optimization path is shown for comparison.}
\label{fig:fB}
\end{figure}

\begin{figure}
\centering
\includegraphics[width=0.6\textwidth]{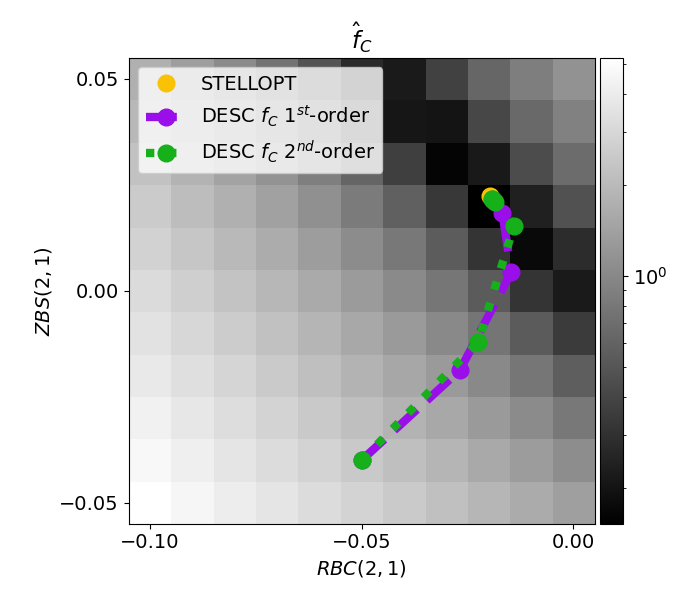}
\caption{Quasi-symmetry optimization paths using the two-term objective in DESC. The optimal STELLOPT solution is shown for comparison.}
\label{fig:fC}
\end{figure}

\begin{figure}
\centering
\includegraphics[width=0.6\textwidth]{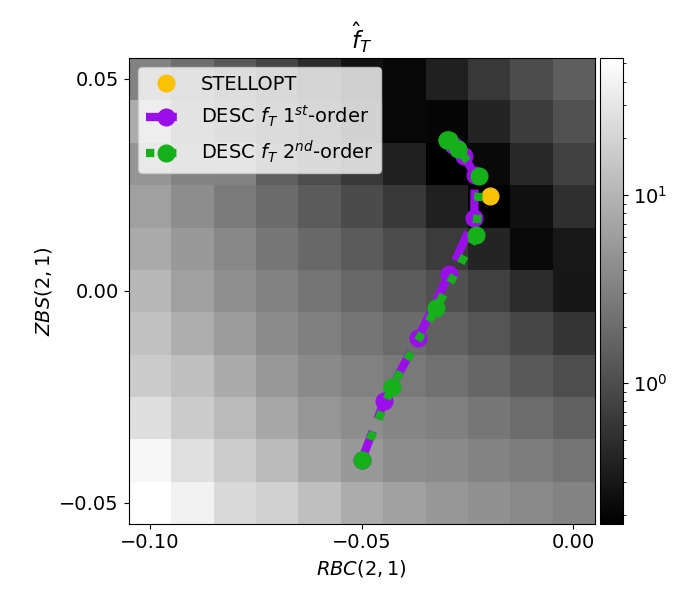}
\caption{Quasi-symmetry optimization paths using the triple product objective in DESC. The optimal STELLOPT solution is shown for comparison.}
\label{fig:fT}
\end{figure}

Figures \ref{fig:fB}, \ref{fig:fC}, and \ref{fig:fT} show the optimization paths of the DESC first and second-order methods for each of the three quasi-symmetry objectives. 
The STELLOPT optimization path for its Boozer form objective is included in Figure \ref{fig:fB}, and the final solution is also indicated in Figures \ref{fig:fC} and \ref{fig:fT} for reference. 
The optimization landscape shown in gray scale in these plots are the respective dimensionless scalar error metrics given in Section \ref{sec:quasi-symmetry}, which are proportional but not equivalent to the least-squares cost functions that are actually being minimized. 
Comparing across all three figures reveals that each of the objective functions has a similar narrow valley of quasi-symmetric solutions, although the global minima are not identical and are still far from perfect quasi-symmetry. 
The DESC optimization runs targeting the Boozer coordinate and two-term objectives converge to results very close to the STELLOPT solution, while the triple product minimum is farther along the valley. 

Although these two optimization codes converge to similar results, their paths to arrive there from the initial configuration are very different. 
The Levenberg-Marquadt algorithm in STELLOPT rarely permits the full Gauss-Newton step in favor of a more conservative approach that is often closer to the steepest descent direction. 
DESC's algorithm is conceptually similar with an adaptive trust-region radius to improve robustness, but allows the full Gauss-Newton step whenever possible. 
As a consequence, DESC converges to the optimum in a few large steps while STELLOPT requires nearly a hundred smaller steps to traverse the optimization landscape in this example (see Figure \ref{fig:fB}). 
In a process analogous to geodesic acceleration in the Levenberg-Marquadt algorithm, the optimization step size in DESC is further increased by including higher-order approximations. 
Contrasting the first and second-order results reveals that the additional derivative information enables the optimizer to take slightly larger steps, which can cause faster convergence in fewer iterations. 

\begin{figure}
\centering
\includegraphics[width=0.6\textwidth]{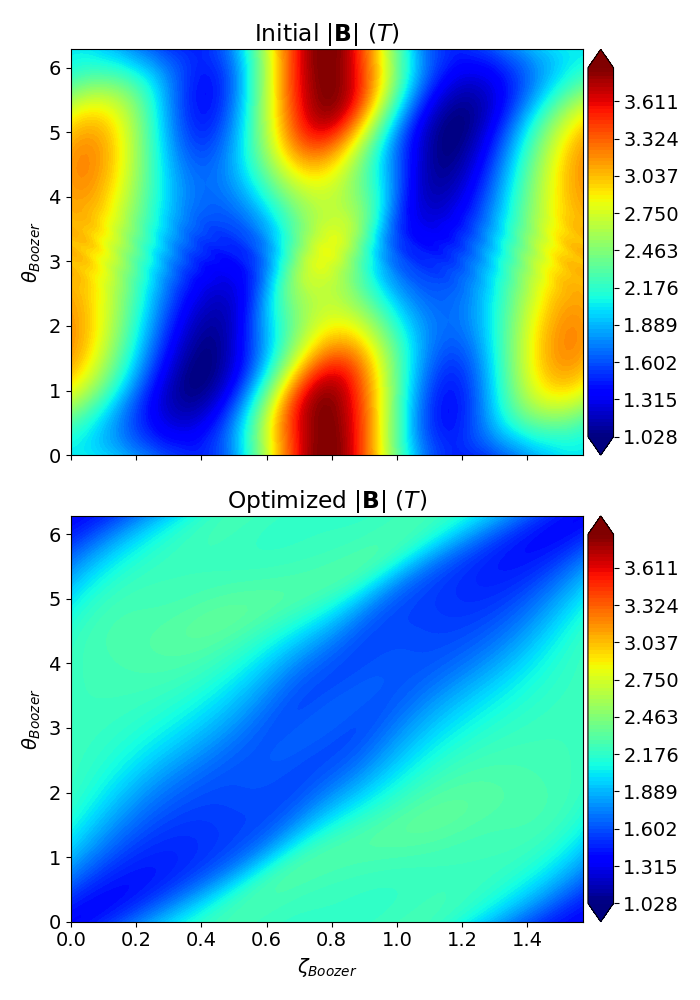}
\caption{Contours of magnetic field magnitude in Boozer coordinates for the initial equilibrium (top) and optimized solution (bottom) targeting the Boozer measure of quasi-symmetry.}
\label{fig:Booz}
\end{figure}

\begin{figure}
\centering
\includegraphics[width=0.6\textwidth]{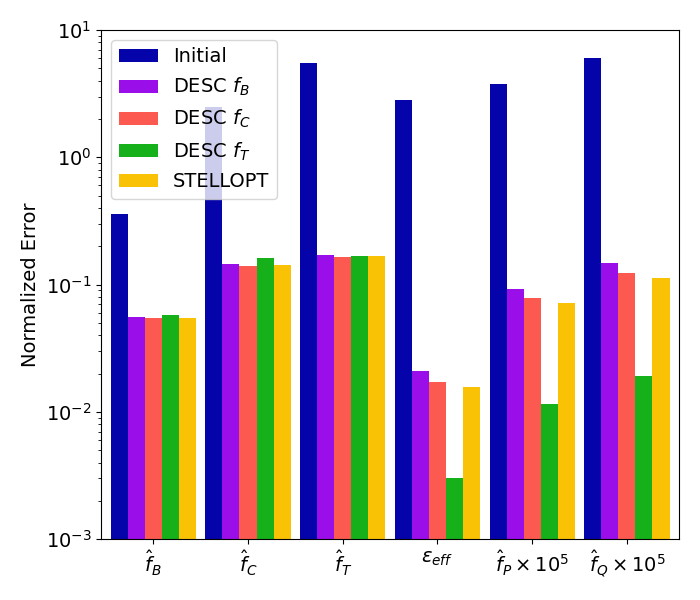}
\caption{Comparison of quasi-symmetry and neoclassical confinement errors for the initial and optimized configurations. Only the second-order DESC results are shown. The three quasi-symmetry metrics were computed in DESC; the effective ripple was computed with NEO from VMEC equilibria; the particle and heat fluxes were computed in SFINCS also using VMEC equilibria.}
\label{fig:errors}
\end{figure}

Figure \ref{fig:Booz} gives a qualitative picture of the quasi-helical symmetry optimization though a comparison of $|B|$ in Boozer coordinates between the initial and final configurations. 
The contours of the optimized solution clearly have the desired slope of $4/1$ that the original stellarator was lacking, although the maximum field strength was also changed in the process. 
However, all of these quasi-symmetry metrics are only used as proxy functions for the true goal of particle confinement. 
In order to validate that the quasi-symmetry optimization does improve the confinement properties, the NEO \citep{Nemov1999} and SFINCS codes \citep{sfincs} were run on the solutions to calculate their neoclassical confinement, and the results for each equilibrium are included in Figure \ref{fig:errors}. 
The quasi-symmetry errors were all computed in DESC, including for the STELLOPT solution which was re-computed with the DESC equilibrium solver. 
The optimal boundary coefficients from each solution were also used to compute high-resolution VMEC equilibria for inputs to NEO and SFINCS, including for the DESC solutions. 
Since the first and second-order results all converged to within $1.3~\text{mm}$ of each other, the errors were only evaluated for the second-order DESC results. 

A surprising result is that all four of the optimized solutions have nearly equivalent values for the three different quasi-symmetry errors, despite each solution targeting a different form during optimization. 
This suggests that the optimization landscape is flat near the global minimum: all of the equilibria in the phase space ``valley'' have similar levels of quasi-symmetry. 
The confinement results tell a different story, however. 
$\epsilon_{eff}$ is the effective ripple in the $1/\nu$ regime and is commonly used as a measure of particle transport \citep{Nemov1999}. 
$\hat{f}_{P}$ and $\hat{f}_{Q}$ are the normalized particle and heat fluxes of the ions: 
\begin{subequations}
\begin{align}
\hat{f}_{P} = \frac{\bar{R}}{\bar{n}\bar{v}} \left\langle \int d^3 v f_{i} \mathbf{v}_{mi} \cdot \nabla\rho \right\rangle \\
\hat{f}_{Q} = \frac{\bar{R}}{\bar{n}\bar{v}^3\bar{m}} \left\langle \int d^3 v f_{i} \frac{m_{i} v^{2}}{2} \mathbf{v}_{mi} \cdot \nabla\rho \right\rangle
\end{align}
\end{subequations}
where $\bar{R} = 1~\text{m}$, $\bar{n} = 1\times10^{20}~\text{m}^{-3}$, $\bar{v} = \sqrt{2 \bar{T} / \bar{m}}$, $\bar{T}=1~\text{keV}$, and $\bar{m}$ is the proton mass. 
All of the optimized solutions substantially reduced the effective ripple and both fluxes, confirming that quasi-symmetry is correlated with neoclassical confinement; but the relationship is complex. 
The DESC optimization targeting the triple product metric, which converged to a result farther away in parameter space from the other solutions, achieved nearly an order of magnitude better confinement than the others according to these measures. 
This lack of correlation supports previous claims that quasi-symmetry is not an accurate indicator of transport levels \citep{Martin2020,Rodriguez2022}. 
The triple product quasi-symmetry measure is the only one that does not depend on equilibrium conditions, and this independence from the force balance constraints could aid the optimization process. 
Unlike the other two metrics which depend on flux surface averaged quantities, the truly local nature of the triple product definition could also explain why it was the most effective proxy for confinement in this optimization example. 

\subsection{Computation Speed}
\label{sec:speed}

Computation times for running realistic quasi-symmetry optimization problems in DESC were benchmarked against STELLOPT. 
STELLOPT is a parallelized FORTRAN code that runs on multiple CPUs, while DESC is a python code aided by just-in-time compilation that is currently compatible with a single CPU or GPU. 
All tests were run on the Traverse computing cluster at Princeton University using AMD EPYC 7281 model CPUs, which have 16 cores, and NVIDIA V100 Tensor Core GPUs, wich have 32G of memory. 
The same numerical resolutions were used between the two codes for a fair comparison. 

\begin{table}
\caption{STELLOPT computation times (in seconds). The VMEC equilibrium inputs used were: \texttt{MPOL = 9}, \texttt{NTOR = 8}, \texttt{NS\_ARRAY = 17 33 65}, and \texttt{FTOL\_ARRAY = 1E-8 1E-10 1E-12}. The BOOZ\_XFORM optimization inputs used were: \texttt{MBOZ = 17}, \texttt{NBOZ = 16}, and \texttt{TARGET\_HELICITY(65) = 0}.}
\label{tab:stellopt}
\begin{tabular}{c|r r r r}
~ & \multicolumn{4}{c}{Number of Optimization Parameters} \\
CPUs & 8 & 24 & 48 & 80 \\
\hline
2 & 362 & 939 & 1757 & 2831 \\
4 & 188 & 399 & 665 & 1074 \\
8 & 150 & 246 & 346 & 550 \\
16 & 118 & 180 & 234 & 322
\end{tabular}
\end{table}

\begin{table}
\caption{DESC computation times (in seconds) for 560 optimization parameters. The numerical resolution used was $L=M=N=8$ for the equilibria and $M=N=16$ for the Boozer spectrum with the $f_{B}$ objective. Quasi-symmetry was targeted on the $\rho=1$ surface.}
\label{tab:desc}
\begin{tabular}{c r|r r r|r r r}
~ & ~ & \multicolumn{3}{c}{Initial Iteration} & \multicolumn{3}{c}{Subsequent Iterations} \\
Objective & ~ & Perturb & Solve & Total & Perturb & Solve & Total \\
\hline
$f_C$ & 1 GPU & 157 & 33 & 190 & 2 & 3 & 5 \\
$f_T$ & 1 GPU & 89 & 33 & 122 & 2 & 3 & 5 \\
$f_B$ & 1 CPU & 763 & 174 & 937 & 409 & 24 & 434 \\
$f_C$ & 1 CPU & 165 & 168 & 333 & 12 & 25 & 37 \\
$f_T$ & 1 CPU & 250 & 169 & 419 & 16 & 23 & 39
\end{tabular}
\end{table}

Table \ref{tab:stellopt} gives the computation times of a single step of the STELLOPT Levenberg-Marquadt optimization algorithm for a range of phase space dimensions and at different levels of parallelization. 
The number of optimization parameters used corresponds to optimizing all of the stellarator-symmetric boundary modes within the resolution $M,N \leq 1, 2, 3, 4$, excluding the major radius. 
Since the bottleneck process of the STELLOPT algorithm is computing gradients through finite differences at each optimization step, its computation time is expected to increase linearly with the number of optimization parameters, and decrease linearly with the number of parallel CPUs that are sharing the workload. 
The timing data in this table confirm that scaling -- the longer run times of larger optimization problems can be reduced through greater parallelization. 
However, for a moderately sized problem on a demanding number of state-of-the-art CPU cores, STELLOPT still takes several minutes per iteration, which results in hours of total computation time. 

In contrast, table \ref{tab:desc} gives the computation times for a single Gauss-Newton trust-region optimization step in DESC, broken down between the perturbation and equilibrium solve involved at each iteration. 
Only the first-order optimization method was tested to give the closest comparison to STELLOPT, but including the second-order terms does not add significantly to the perturbation times as explained previously. 
The equilibrium solve time is heavily dependent on the number of Gauss-Newton steps required to solve (\ref{eq:fmin}) to the desired tolerances. 
The average times shown in this table are for ten iterations of the equilibrium subproblem, but in practice the number will vary between optimizer iterations (larger optimization steps typically require more force balance correction). 
Thanks to the use of automatic differentiation, the computation times are independent of the number of optimization parameters -- these results included all $560$ boundary modes in the $M, N \leq 8$ equilibrium resolution. 
Because the objective functions get compiled to work with automatic differentiation the first time they are called, the initial iteration takes substantially longer than subsequent ones. 
The benefit is that calls to the compiled functions can then be evaluated much faster, yielding large savings in the long run when many optimization steps are required. 
These functions only need to be recompiled if the resolution of the equilibrium changes, which typically does not occur in a traditional optimization procedure. 
Even with a high level of parallelization in STELLOPT, the reliance on finite differencing prevents it from approaching the speed of automatic differentiation for large numbers of optimization parameters. 

This table also reveals a large speedup of running on a GPU: the compile times are roughly halved and iterations are about $8$ times faster than on the CPU for this problem. 
It is important to note that the same code can run on either a CPU or GPU without any modifications required by the user. 
Special attention is needed for discussion of the quasi-symmetry objective $f_{B}$: because the conversion to Boozer coordinates requires re-evaluating Fourier basis functions in the new coordinates, differentiating through this function with automatic differentiation is very memory intensive. 
The numerical resolution used in these tests was too large to run the Boozer form on the GPU with limited memory capacity. 
The perturbation time of the Boozer form on the CPU is also much slower than the other quasi-symmetry objectives, but it is still a significant improvement over STELLOPT especially when considering the number of CPUs and optimization parameters involved. 
Whether running on a single CPU or GPU, the total run time is orders of magnitude faster using DESC compared to STELLOPT. 

\section{Conclusions}

DESC is designed to be the next-generation stellarator optimization code. 
Its unique architecture allows it to take advantage of automatic differentiation and seamlessly pass that derivative information from its equilibrium solver to the optimizer, resulting in dramatic speed improvements over existing tools. 
This paper has explained the DESC optimization approach and demonstrated the simplicity of using the python code. 
Through examples of quasi-symmetry optimization, it has shown that DESC can achieve similar if not better quality results as conventional tools such as STELLOPT in orders of magnitude less computation time, and even other modern tools like SIMSOPT are not designed for this level of speed performance. 
The triple product metric is believed to be the best quasi-symmetry objective due to its lack of physics assumptions and its numerical advantages, but more test cases are needed to confirm if this result generalizes to other problems. 
The purpose has been to showcase the potential of DESC, and discovering reactor-relevant stellarator designs is left for future work. 

As an open source software with a modular structure, the DESC code suite is continually improving upon existing performance and expanding to provide new functionality. 
The example quasi-symmetry optimizations used default options, but there is great flexibility in specifying alternatives to better accommodate certain problems. 
Quasi-symmetry was provided as an example optimization objective due to its popularity in the literature, but other physics and engineering objectives should be added and used in tandem. 
Thanks to automatic differentiation, the process of incorporating new optimization targets with exact derivative information is relatively simple for the developer. 
There is also room to add alternatives to the Gauss-Newton trust-region optimization method, such as global optimization algorithms. 
More work is needed to better understand the trade-offs between robustness and convergence speed among the various possibilities. 
Plans to extend DESC to handle free-boundary equilibria and single-stage coil optimization are already under development. 
Anyone who is interested in taking advantage of these capabilities is strongly encouraged to install and use the publicly available code, and recommendations for additional features are always welcome. 

Due to its efficient computations, DESC is well positioned to perform a large-scale exploration of the stellarator design space. 
The perturbation and continuation methods could provide physical insight on the bifurcation of tokamaks into quasi-axisymmetric stellarators \citep{Plunk2020}. 
DESC could also be used to generate a large database of equilibria that could then be analyzed with the aid of machine learning techniques to identify regions in the parameter landscape that deserve closer attention. 
Another important application is equilibrium reconstruction, which is an optimization problem to find the equilibrium parameters that best match experimental diagnostics. 
The rate of generating full equilibria reconstructions with conventional tools is disappointingly slow \citep{Hanson2009}, but DESC could be used to provide timely analysis of experiments. 
Whether improving the usefulness of existing devices or discovering better designs for future reactors, the optimization capabilities of DESC open new possibilities for stellarator research that were previously unavailable. 

Thank you to Eduardo Rodriguez for many insightful conversations about the nature of quasi-symmetry and its potential for stellarator optimization, and to Albert Mollen and Elizabeth Paul for their instruction on the SFINCS code. 
Thank you to the SIMSOPT development team, whose python version of BOOZ\_XFORM was the basis for the implementation in DESC. 

This work was supported by the U.S. Department of Energy under contract numbers DE-AC02-09CH11466,
DE-SC0022005 and Field Work Proposal No. 1019. 
The United States Government retains a non-exclusive, paid-up, irrevocable, world-wide license to publish or reproduce the published form of this manuscript, or allow others to do so, for United States Government purposes. 

The data that support the findings of this study are available from the corresponding author upon reasonable request. 
The latest version of the DESC code can be accessed at the public repository: \url{https://github.com/PlasmaControl/DESC}. 
The results in this paper were generated from the git commit hash \texttt{0fcc708}, and the input files used are available with the publication. 

\bibliographystyle{jpp}
\bibliography{bibliography.bib}

\end{document}